\documentclass[twoside,a4paper,11pt]{sea10}
\usepackage{graphicx}
\usepackage{hyperref}
\usepackage{movie15}
\usepackage{natbib}
\begin{document}
\pagenumbering{arabic}
\pagestyle{myheadings}
\thispagestyle{empty}
{\flushleft\includegraphics[width=\textwidth,bb=58 650 590 680]{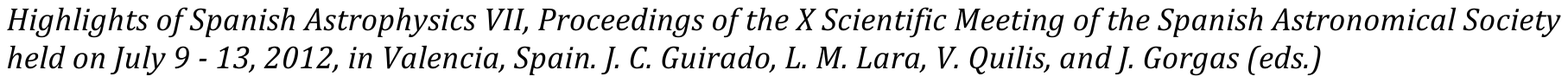}}
\vspace*{0.2cm}
\begin{flushleft}
{\bf {\LARGE
%
A multipurpose 3-D grid of stellar models
%
}\\
\vspace*{1cm}
%
J. Ma\'{\i}z Apell\'aniz$^{1}$
%
}\\
\vspace*{0.5cm}
%
$^{1}$
Instituto de Astrof\'{\i}sica de Andaluc\'{\i}a-CSIC, Glorieta de la Astronom\'{i}a s/n, \linebreak 18008 Granada, Spain
%
\end{flushleft}
%
\markboth{
A multipurpose 3-D grid of stellar models
}{ 
%
Ma\'{\i}z Apell\'aniz
%
}
\thispagestyle{empty}
\vspace*{0.4cm}
\begin{minipage}[l]{0.09\textwidth}
\ 
\end{minipage}
\begin{minipage}[r]{0.9\textwidth}
\vspace{1cm}
\section*{Abstract}{\small
%
The last two decades have produced a proliferation of stellar atmosphere grids, evolutionary tracks, and isochrones which 
are available to the astronomical community from different internet services. However, it is not straightforward (at least 
for an inexperienced user) to manipulate those models to answer questions of the type: What is the spectral energy 
distribution of a 9000 K giant? What about its $J$-band magnitude for different metallicities? What can I tell about the mass 
of a star if I know that its unreddened $B-V$ color is $-$0.05 and its luminosity in solar units is $10^5$? The answers to 
those questions are indeed in the models but a series of transformations and combinations involving different variables and 
models are required to obtain them. To make the available knowledge more user friendly, I have combined a number of 
state-of-the-art sources to create a 3-D (effective temperature, luminosity, and metallicity) grid of stellar models for which
I provide calibrated SEDs and magnitudes as well as auxiliary variables such as mass and age. Furthermore, I have generated a 
grid of extinguished magnitudes using the recent Ma\'{\i}z Apell\'aniz et al. (2012) extinction laws and incorporated them 
into the Bayesian code CHORIZOS (Ma\'{\i}z Apell\'aniz 2004).
%
\normalsize}
\end{minipage}
%
%
%
%

\begin{figure}
\centerline{
\includegraphics[width=1.2\linewidth, bb=28 28 566 566]{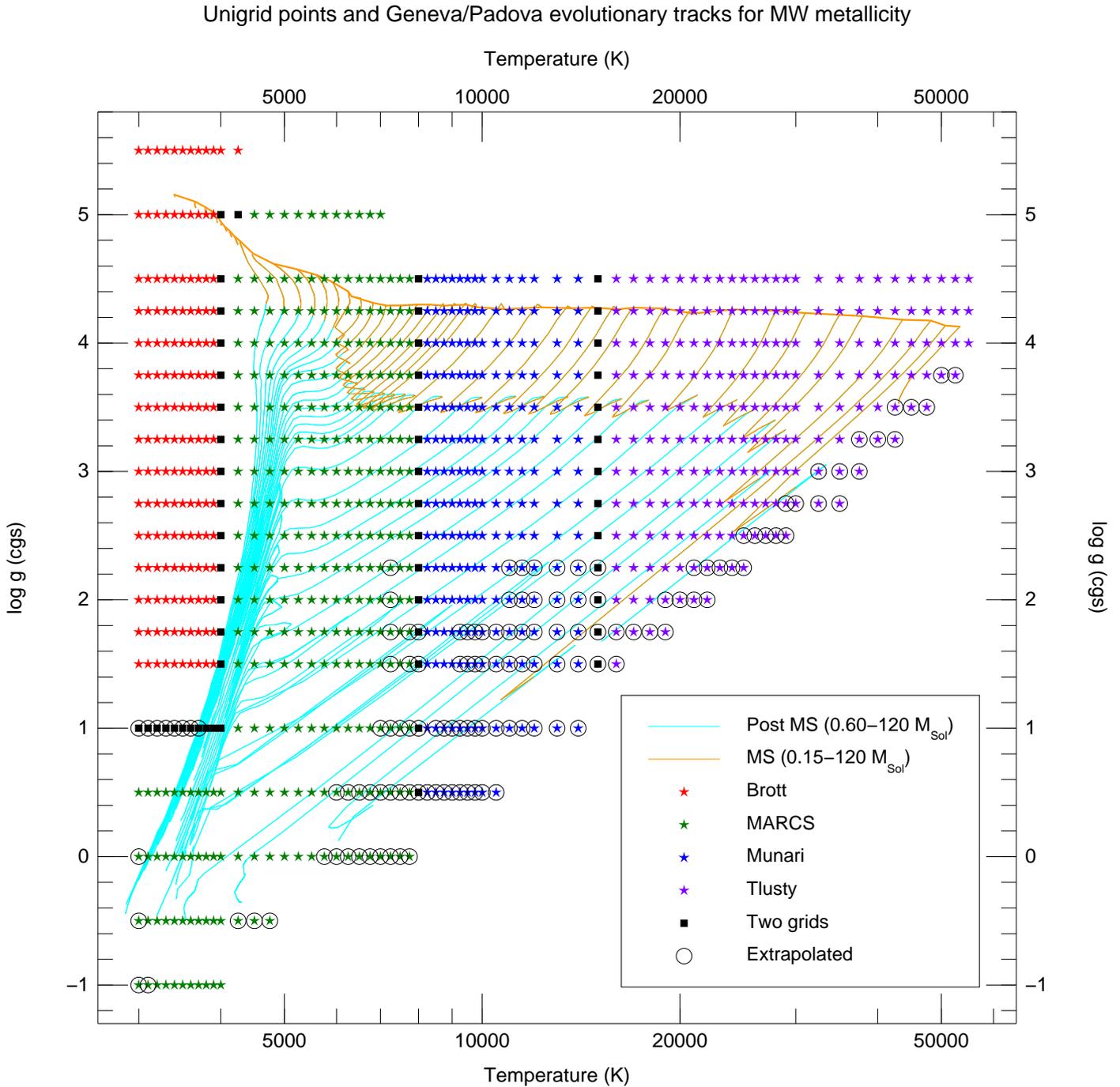}
}
\caption{Grid coverage in the $T_{\rm eff} - \log g$ plane for the MW metallicity. Different colors and symbols are used to indicate the source 
of the used SED at each point, with solid black squares for the cases where the average of two SED families were used. Empty circles indicate those 
points where the SED was extrapolated (i.e. it did not exist in the original grid and it could not be interpolated): those SEDs should be
considered of accuracy than the rest. The Geneva evolutionary tracks for $Z = 0.020$ with standard mass-loss rates 
($9$~M$_\odot \le m_i \le 120 $~M$_\odot$) and the lower Padova evolutionary tracks for $Z = 0.019$ 
($0.15$~M$_\odot \le m_i \le 7$~M$_\odot$) are also plotted.}
\end{figure}

\section{Questions every stellar astronomer has to answer from time to time}

\begin{itemize}
 \item What is the UV-optical-NIR spectral energy distribution of a solar-metallicity 9000 K main-sequence star at a distance of 
     700 pc and an $A_V$ of 1.3 magnitudes?
 \item What about its JHK magnitudes as a function of metallicity?
 \item A star with LMC metallicity has an unreddened $B-V$ color of $-$0.05 and its luminosity is $10^5$ L$_\odot$. How well
     constrained is its mass from that information?
\end{itemize}

\section{Possible sources of answers}

\begin{itemize}
 \item Textbooks or textbook-like references (e.g. \citealt{CarrOstl96}):
 \begin{itemize}
  \item Information may be outdated and/or incomplete. 
  \item Sources are sometimes unclear.
 \end{itemize}
 \item Modern references and websites with stellar atmospheres, evolutionary tracks, and others:
 \begin{itemize}
  \item But each source provides only part of the puzzle: $T_{\rm eff} + \log g$ but not luminosity, photometry but not mass\ldots
  \item Different sources cover only partial ranges in $T_{\rm eff}$, mass, $L$\ldots What about consistency over the full range?
  \item Photometry is usually limited to one or a few systems.
 \end{itemize}
\end{itemize}

\section{What do we want?}

\begin{itemize}
 \item A 3-D cartesian grid with coordinates: 
 \begin{itemize}
  \item $T_{\rm eff}$.
  \item Luminosity class (LC): A parameter equivalent to the spectroscopic one that ranges from 0 (hypergiants) to 5.5 (ZAMS).
  \item $Z$.
 \end{itemize}
 \item What should go at each point in the grid?
 \begin{itemize}
  \item Distance-calibrated SEDs at least from 1000 \AA\ to 2.5 $\mu$m.
  \item $\log L$.
  \item $\log g$.
  \item $m_i$ (mean initial mass).
  \item $\sigma_{m_i}$ (range of initial masses).
  \item $m$ (mean current mass).
  \item Evolutionary phase.
  \item Mean age.
  \item Time spent at a given point (cell).
  \item IMF + age range weight.
 \end{itemize}
 \item Plus photometry:
 \begin{itemize}
  \item Magnitudes, colors, indices, bolometric corrections\ldots
  \item \ldots which require filter throughputs, zero points, and extinction laws.
 \end{itemize}
\end{itemize}

\begin{figure}
\centerline{
\includegraphics[width=1.2\linewidth, bb=28 28 566 566]{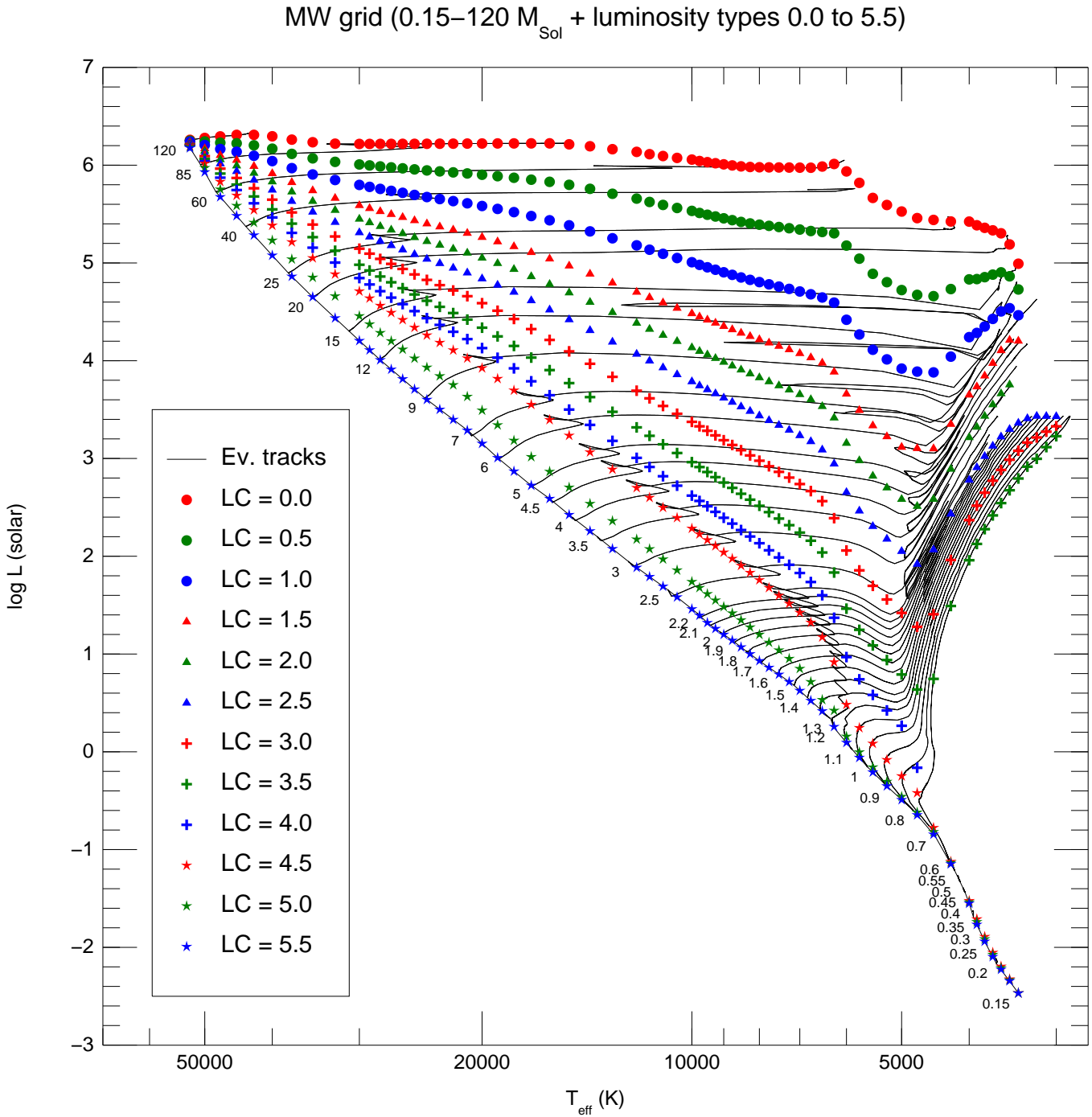}
}
\caption{Building the MW metallicity grid. The black lines are the Geneva/Padova evolutionary tracks between $0.15$ M$_\odot$ and $120$ M$_\odot$ (a label
at the beginning of the track shows the initial mass). Different symbols are used for the luminosity types 0.0\ldots 5.5. Note that luminosity types are 
defined in the grid at 0.1 intervals but only those at 0.5 intervals are shown for clarity.}
\end{figure}

\begin{figure}
\centerline{
\includegraphics[width=1.2\linewidth, bb=28 28 566 566]{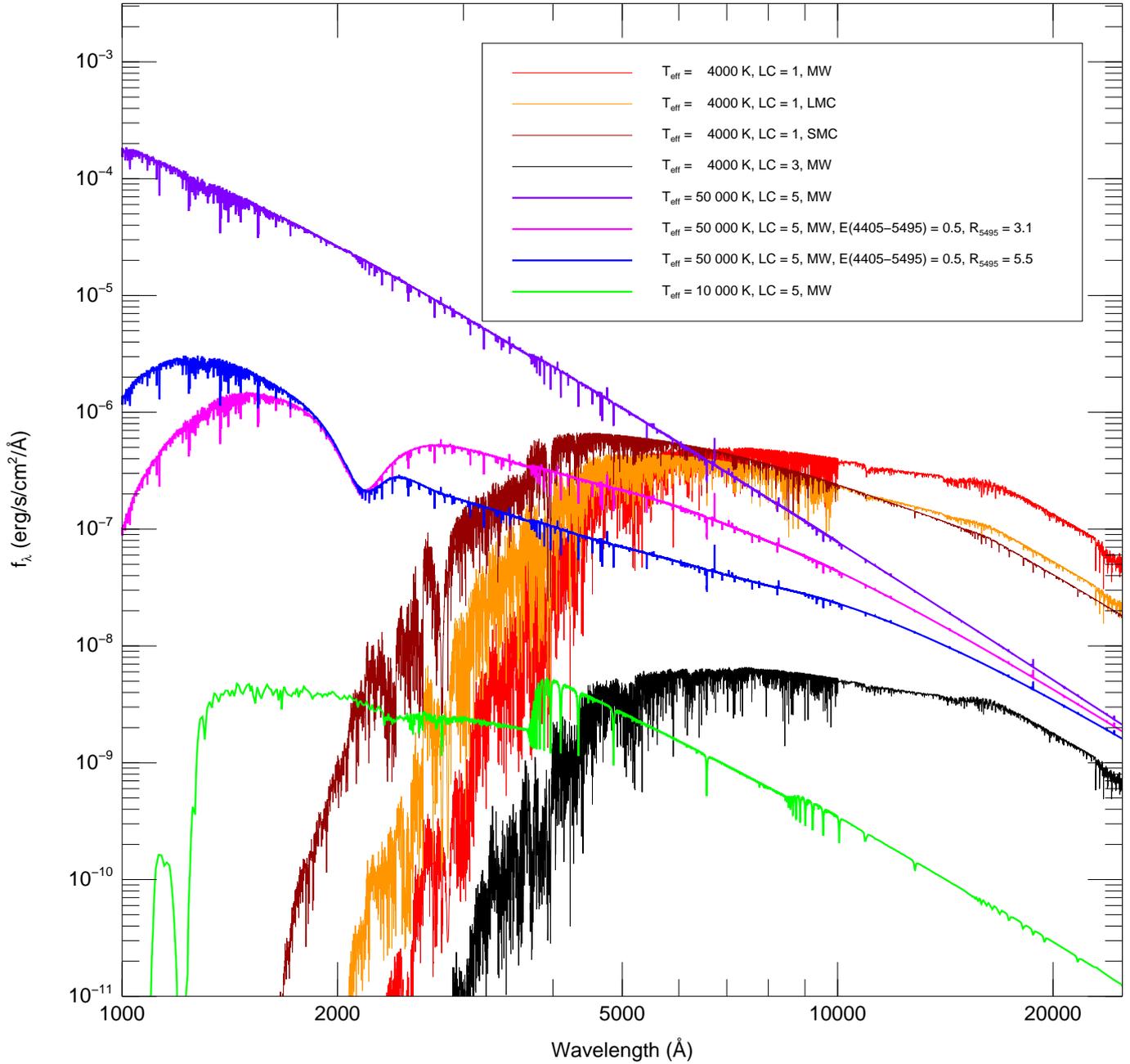}
}
\caption{Sample SEDs from the grid. Fluxes shown for a distance of 10 pc.}
\end{figure}

\section{What do I use? Stellar atmospheres}

\begin{itemize}
 \item Four regions in the $T_{\rm eff} + \log g$ plane for three metallicities (MW, LMC, and SMC) (Fig.~1):
 \begin{itemize}
  \item TLUSTY \citep{LanzHube03,LanzHube07}.
  \item Munari \citep{Munaetal05}.
  \item MARCS \citep{Gustetal75,Gustetal03,Plezetal92}. 
  \item Brott \citep{BrotHaus05}.
 \end{itemize}
 \item Interpolations checked at boundaries to avoid jumps.
 \item Some extrapolations into low gravity are needed due to the lack of publicly existing models at the present time.
\end{itemize}

\section{What do I use? Evolutionary tracks + luminosity classes}

\begin{itemize}
 \item Two mass ranges:
 \begin{itemize}
  \item Geneva \citep{LejeScha01}: $m_i \ge 9 $ M$_\odot$.
  \item Padova \citep{Giraetal00,Salaetal00}: $m_i \le 7$ M$_\odot$.
 \end{itemize}
 \item Luminosity classes (Fig. 2):
 \begin{itemize}
  \item Follow spectroscopic equivalents approximately.
  \item Top [LC=0.0], left [$T_{\rm eff} = $Max$(T_{\rm eff})$], and bottom [LC=5.5] edges of the grid are constant.
  \item Right edge is irregular and metallicity-dependent (Figs.~4-9).
 \end{itemize}
\end{itemize}

\section{What do I use? Photometry and extinction}

\begin{itemize}
 \item Photometry: jmasynphot.
 \begin{itemize}
  \item Written in IDL.
  \item Developed for CHORIZOS \cite{Maiz04c}.
  \item 145 different filters (Johnson-Cousins, Str\"omgren, HST, SDSS, IPHAS, Galex, Tycho-2).
  \item Self-consistent calibration \citep{Maiz05b,Maiz06a,Maiz07a}.
  \item Bolometric corrections calculated.
 \end{itemize}
 \item Extinction: Talk at this meeting by J. Ma\'{\i}z Apell\'aniz.
 \begin{itemize}
  \item Single-parameter ($R_{5495}$) family of extinction laws.
  \item Similar to \citet{Cardetal89} but with lower residuals.
 \end{itemize}
\end{itemize}

\begin{figure}
\centerline{
\includegraphics[width=0.85\linewidth, bb=28 28 566 324]{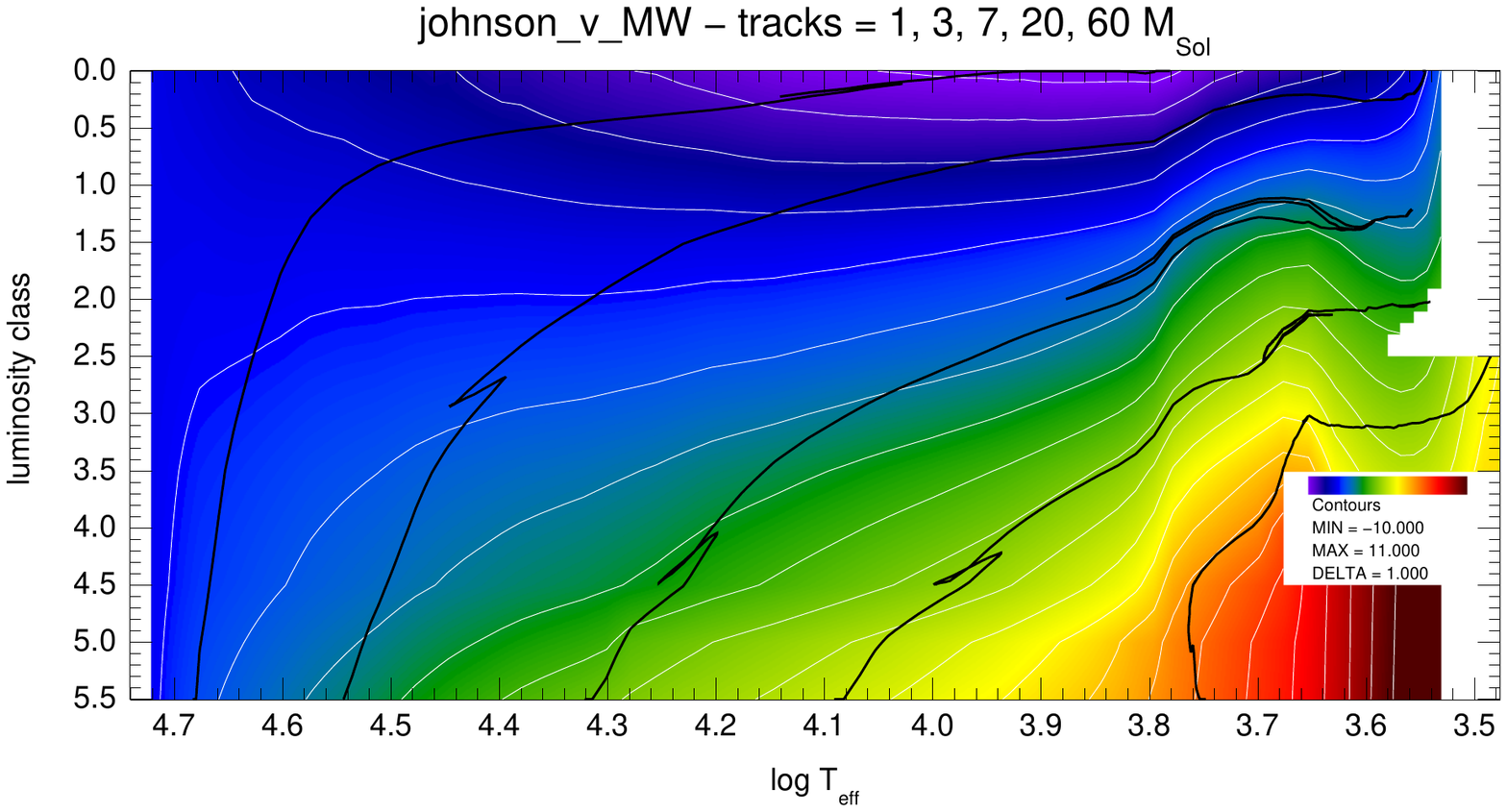}
}
\centerline{
\includegraphics[width=0.85\linewidth, bb=28 28 566 324]{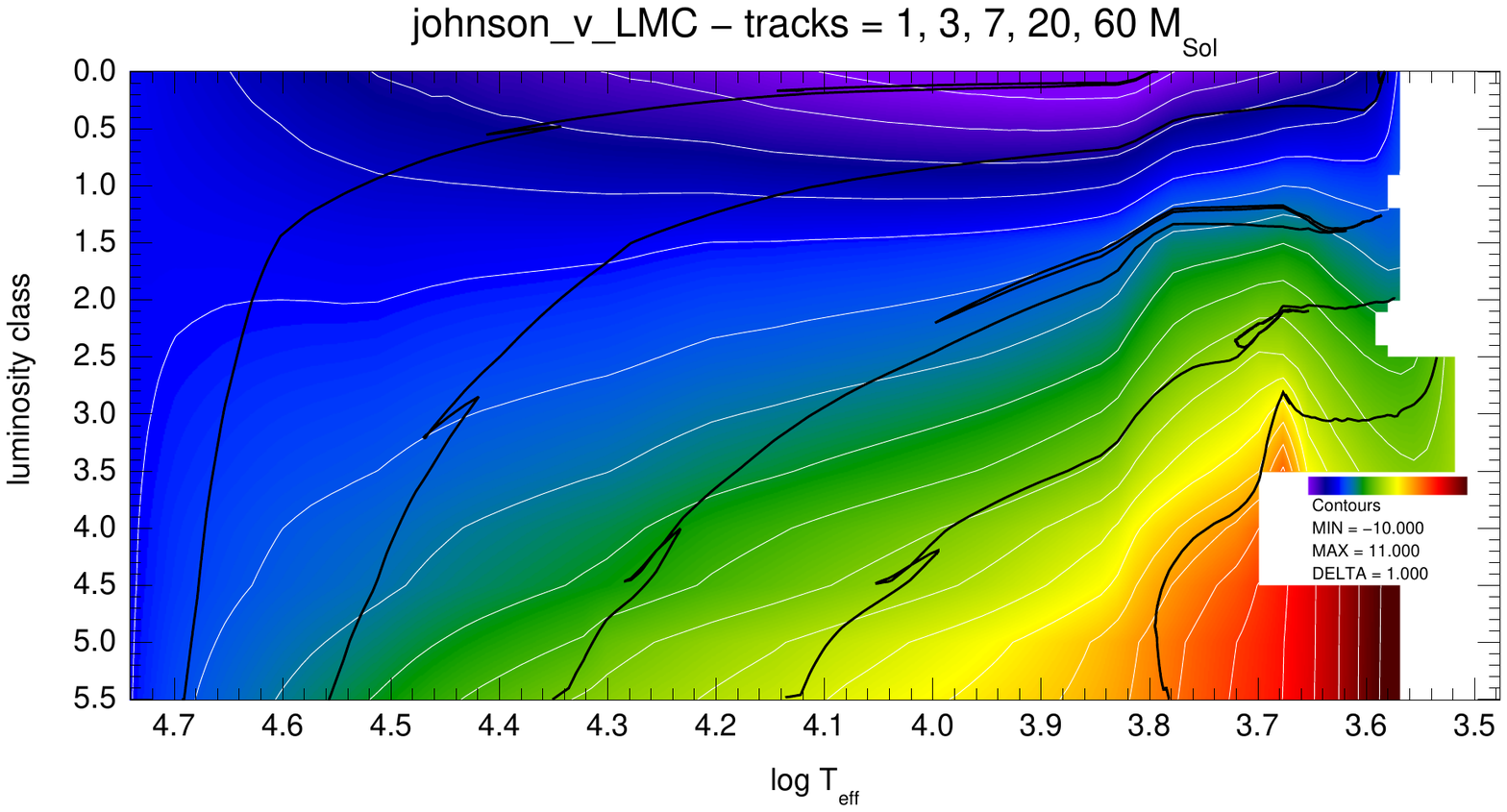}
}
\centerline{
\includegraphics[width=0.85\linewidth, bb=28 28 566 324]{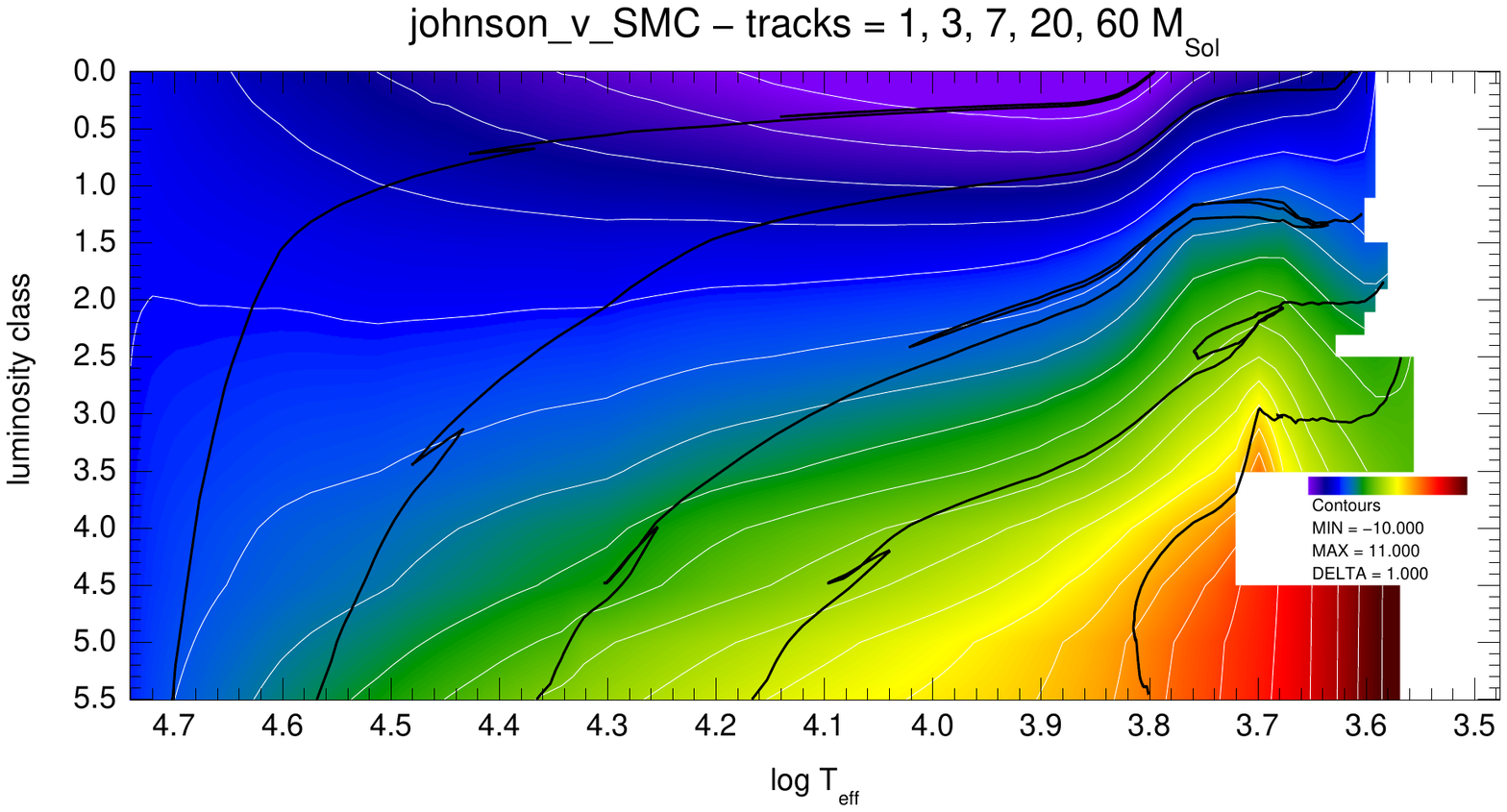}
}
\caption{Johnson $V$ magnitude for the MW (top), LMC (center), and SMC (bottom) $T_{\rm eff}$ - LC grids. In all cases the contours are spaced at 1 
magnitude intervals between $V = 11.0$ and $V = −10.0$. The evolutionary tracks for $m_i$ = 1, 3, 7, 20, and 60 M$_\odot$ are also plotted.}
\end{figure}

\begin{figure}
\centerline{
\includegraphics[width=1.2\linewidth, bb=28 28 566 324]{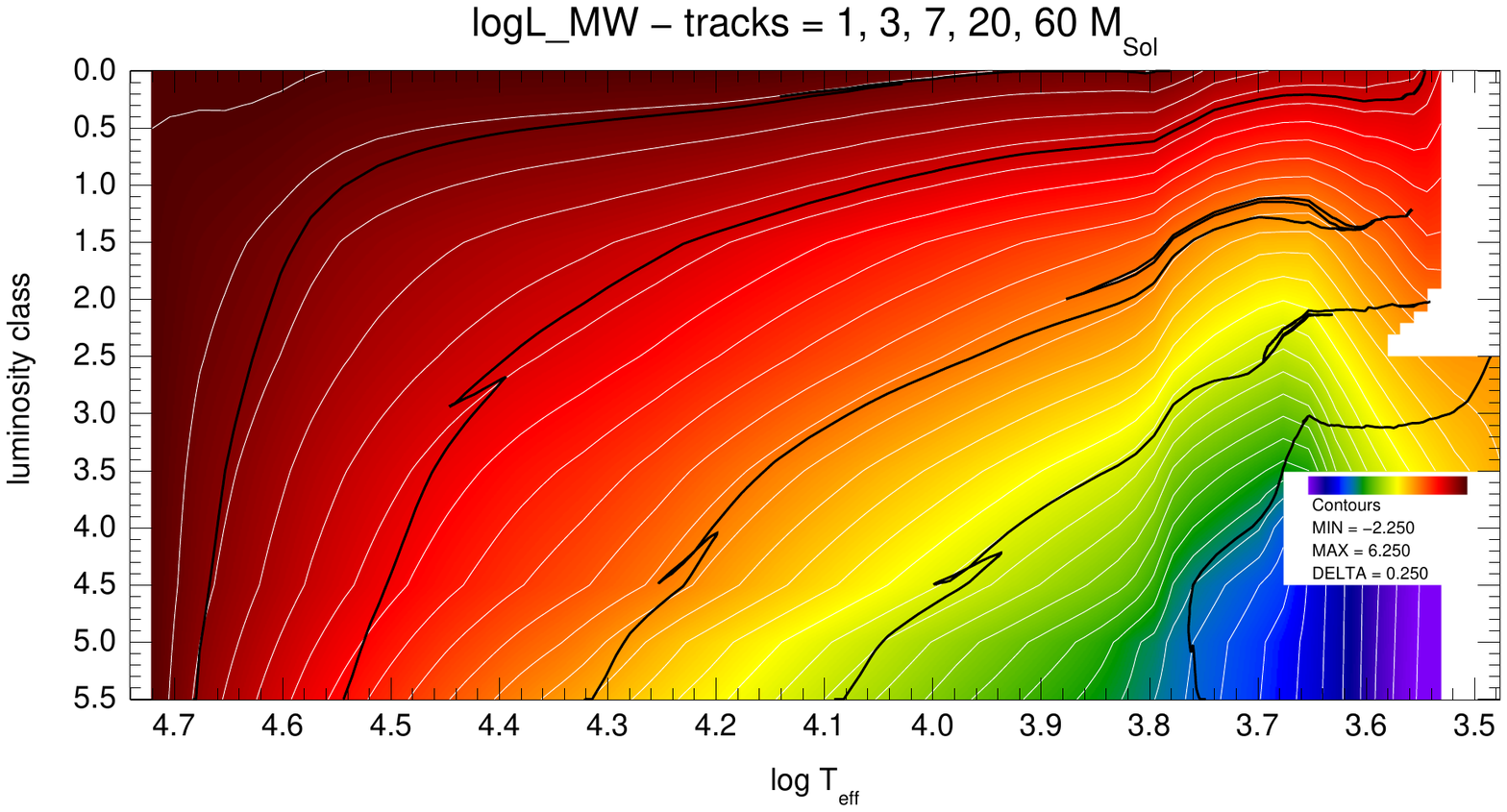}
}
\centerline{
\includegraphics[width=1.2\linewidth, bb=28 28 566 324]{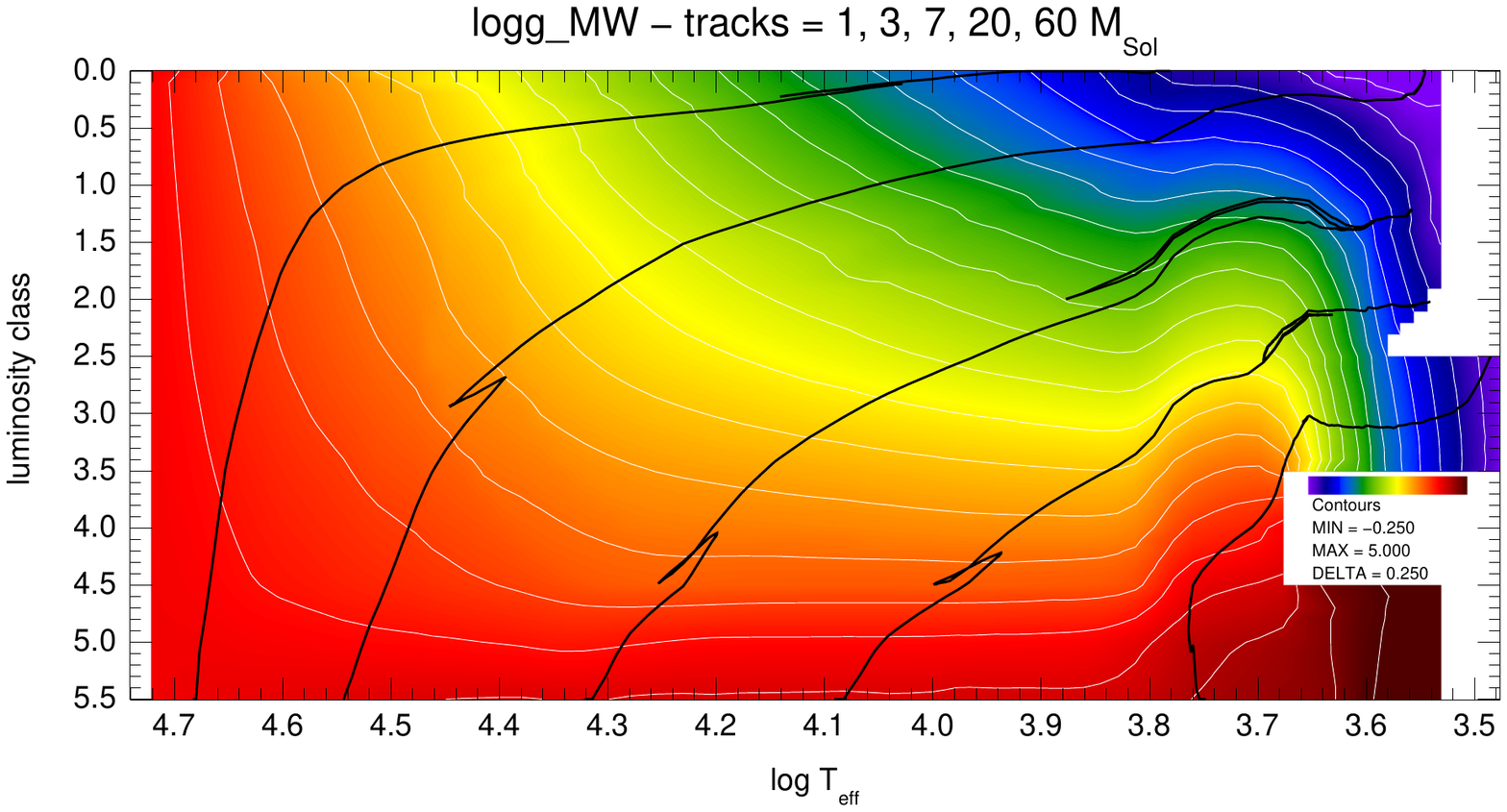}
}
\caption{$\log L$ (top) and $\log g$ (bottom) $T_{\rm eff}$ - LC grid for the MW. The $\log L$ contours range from $-2.25$ to $6.25$ (solar units) 
with a spacing of $0.25$. The $\log g$ contours range from $-0.25$ to $5.00$ (cgs) with a spacing of $0.25$.
The evolutionary tracks for $m_i$ = 1, 3, 7, 20, and 60 M$_\odot$ are also plotted.}
\end{figure}

\begin{figure}
\centerline{
\includegraphics[width=1.2\linewidth, bb=28 28 566 324]{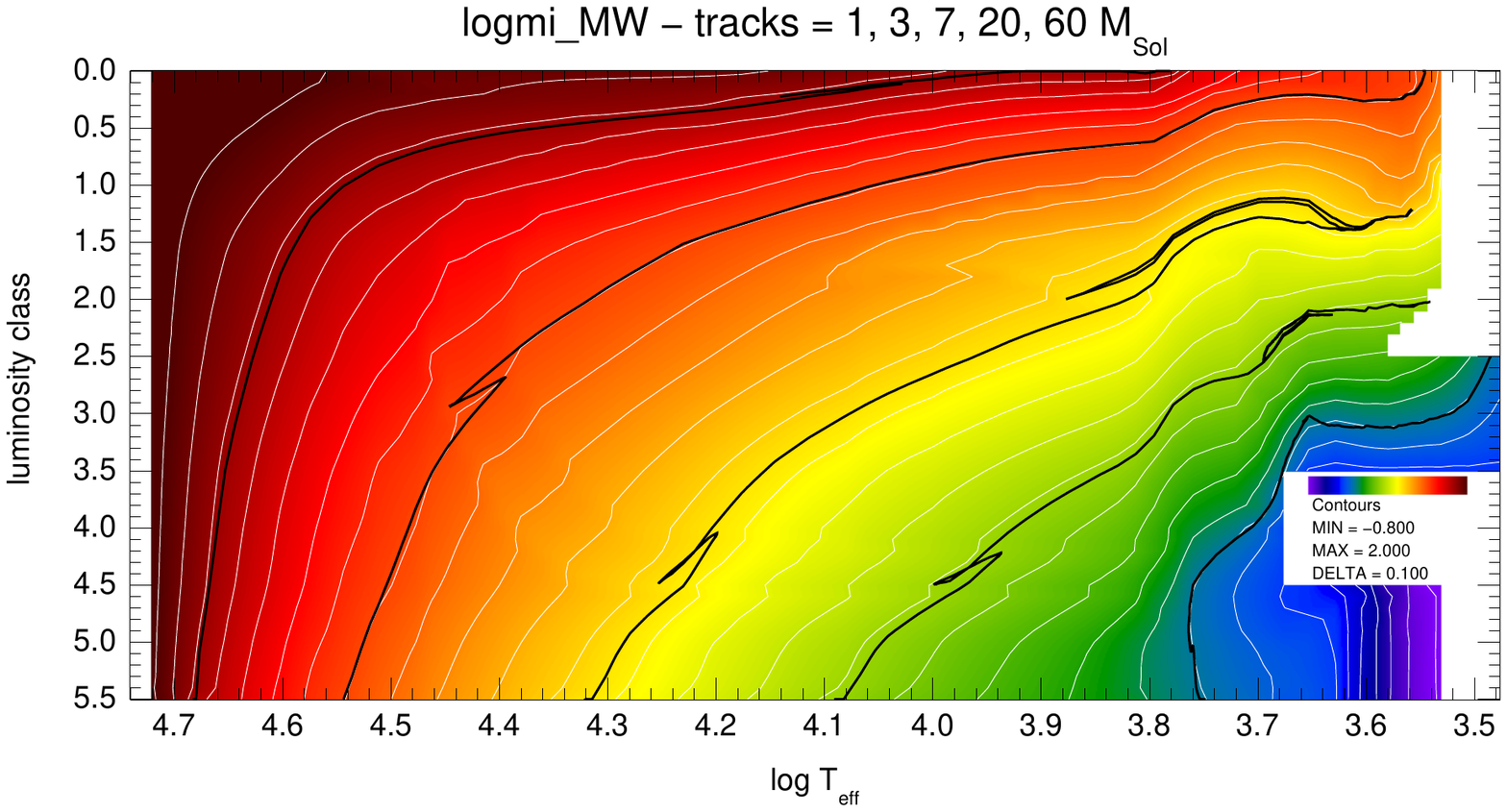}
}
\centerline{
\includegraphics[width=1.2\linewidth, bb=28 28 566 324]{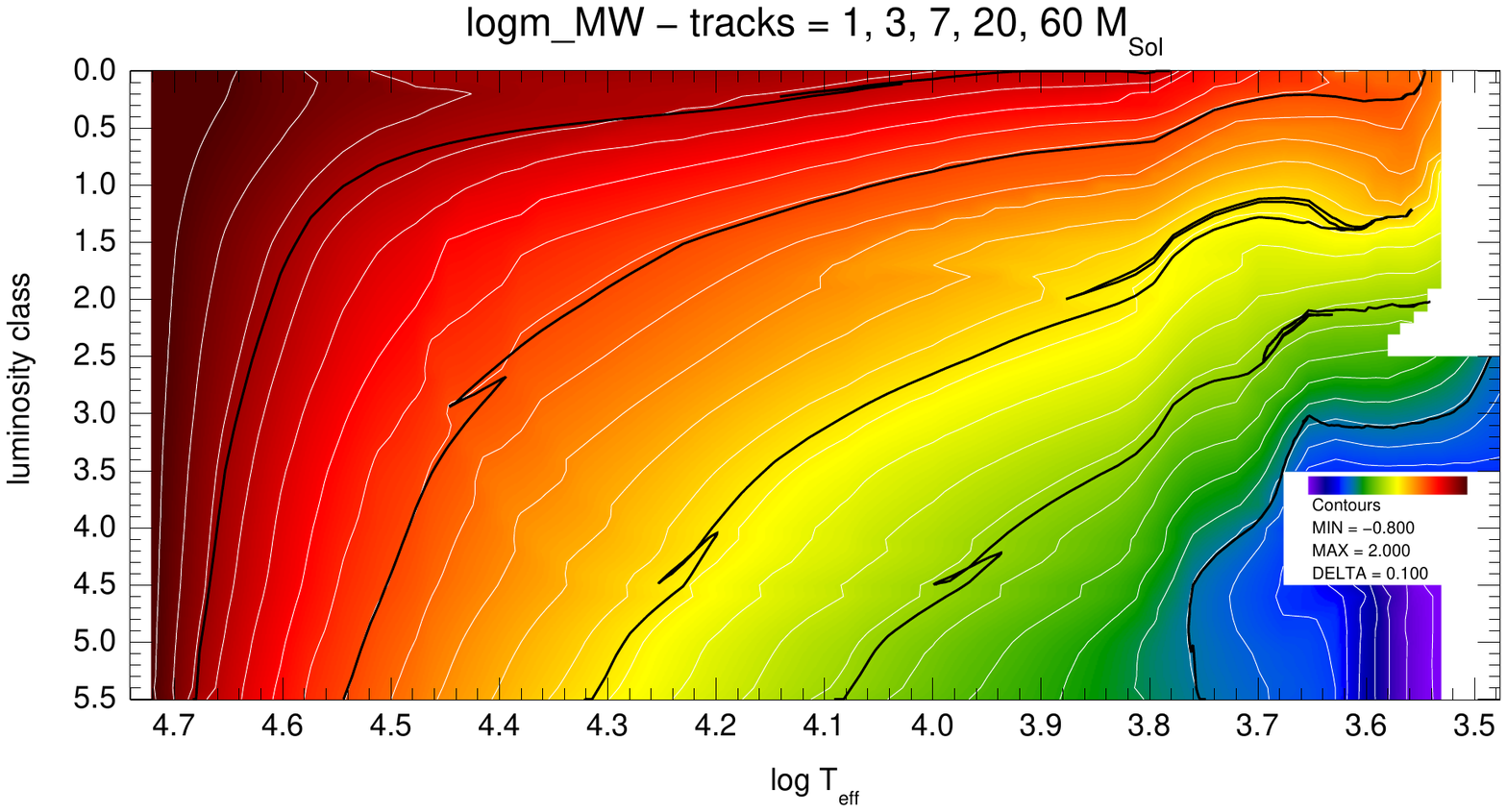}
}
\caption{$\log m_i$ (top) and $\log m$ (bottom) $T_{\rm eff}$ - LC grid for the MW. The contours range from $-0.8$ to $2.0$ (solar units) 
with a spacing of $0.1$. 
The evolutionary tracks for $m_i$ = 1, 3, 7, 20, and 60 M$_\odot$ are also plotted.}
\end{figure}

\section{Applications}

\begin{itemize}
 \item Reference and education.
 \item Filter selection for observational projects.
 \item Exposure time calculators.
 \item Input for CHORIZOS \citep{Maiz04c}:
 \begin{itemize}
  \item Bayesian photometric code: you give me photometry, I give you which SEDs are compatible with it.
  \item New version (3.x) has distance as independent parameter and needs SEDs with absolute flux calibration.
  \item Processing of multi-filter Galactic photometry: from magnitudes to distances, extinction, and physical properties.
 \end{itemize}
\end{itemize}

\section{What is currently not there but may be in the future}

\begin{itemize}
 \item More recent evolutionary tracks with finer grids.
 \item Rotation, stellar winds...
 \item Late evolutionary phases: Wolf-Rayet, post-He flash, white dwarfs...
 \item More metallicity points.
 \item Extension to the MIR.
 \item Interested in collaborating? Contact me!
\end{itemize}

\begin{figure}
\centerline{
\includegraphics[width=0.85\linewidth, bb=28 28 566 324]{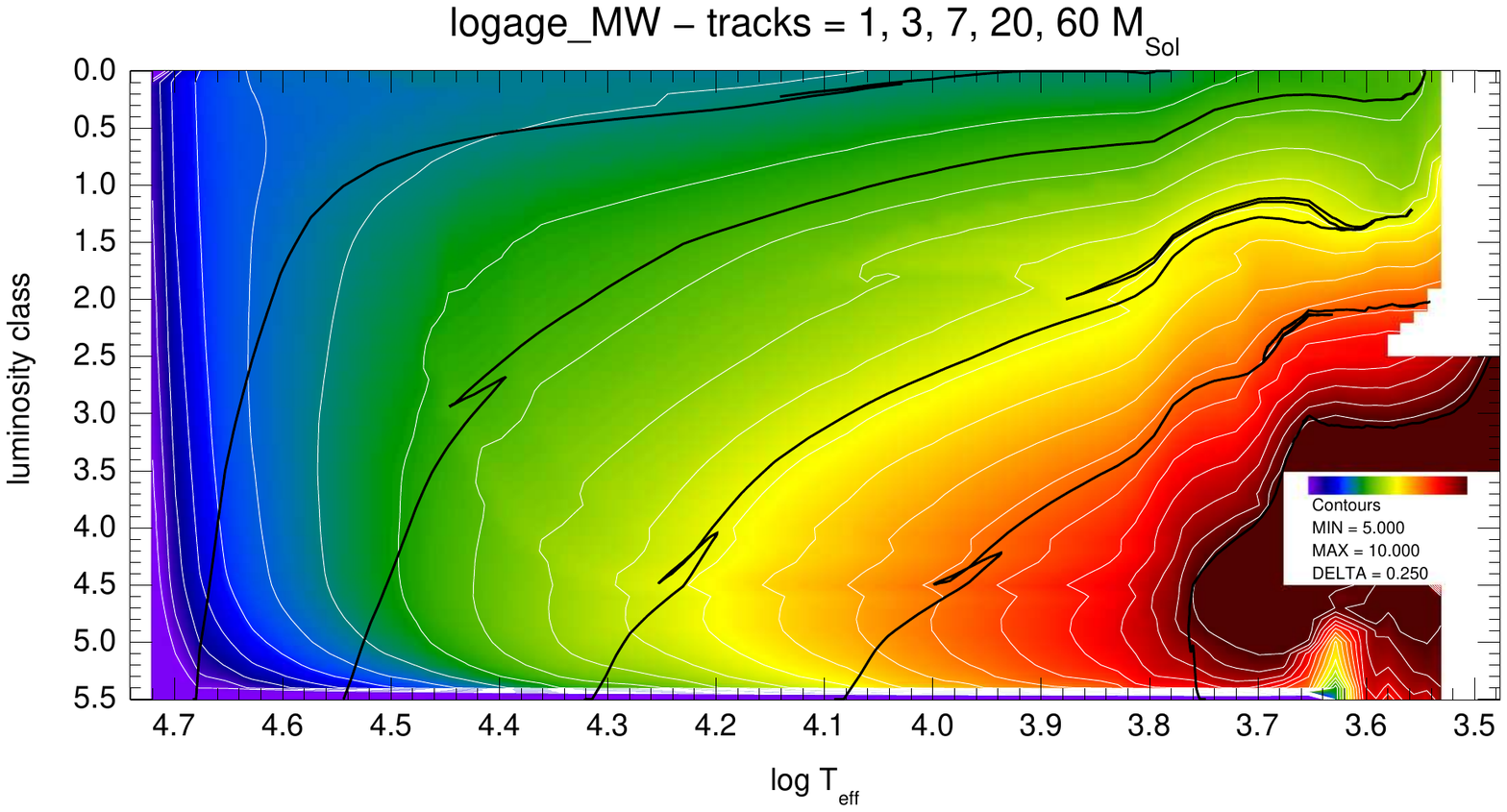}
}
\centerline{
\includegraphics[width=0.85\linewidth, bb=28 28 566 324]{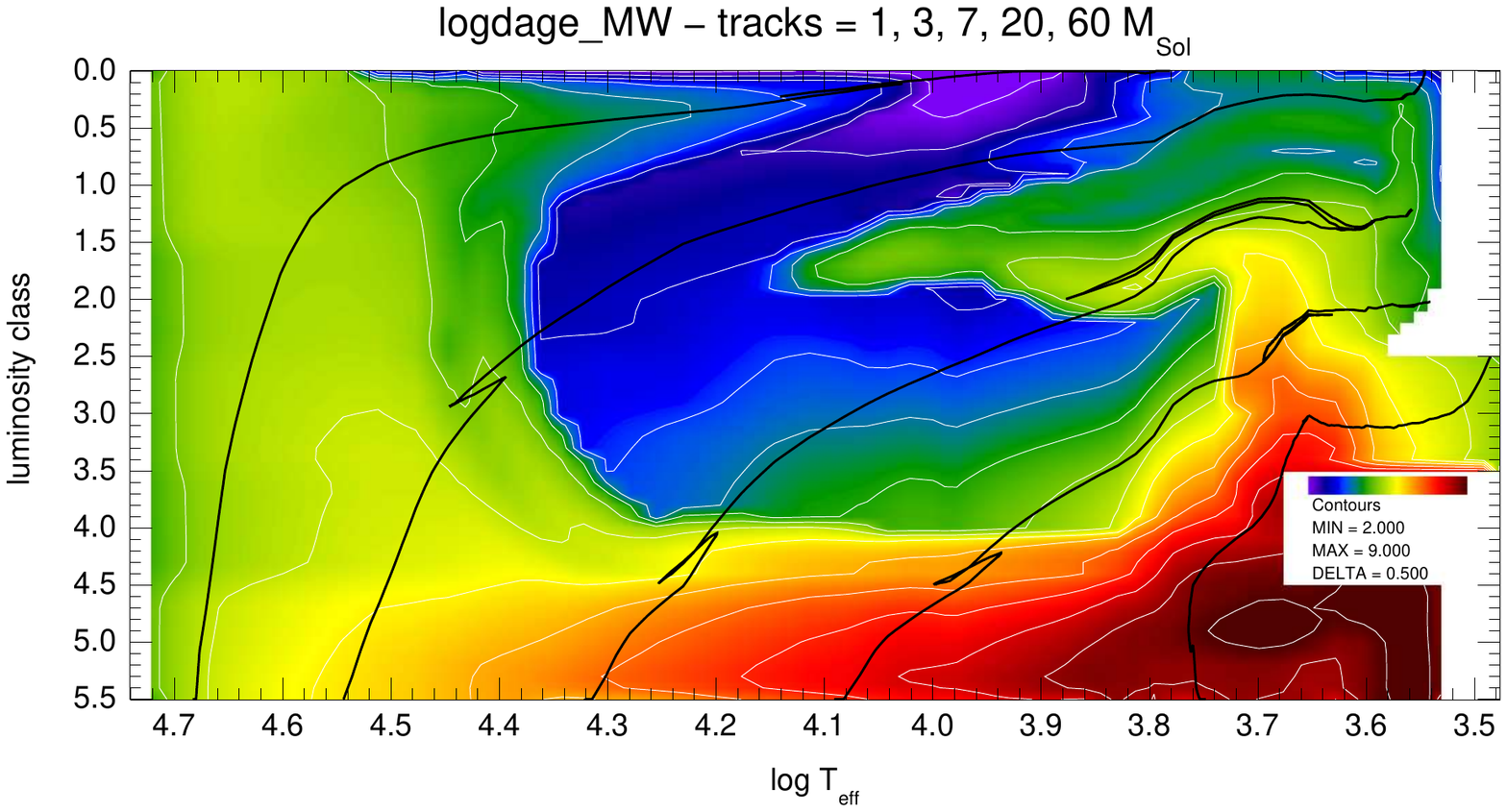}
}
\centerline{
\includegraphics[width=0.85\linewidth, bb=28 28 566 324]{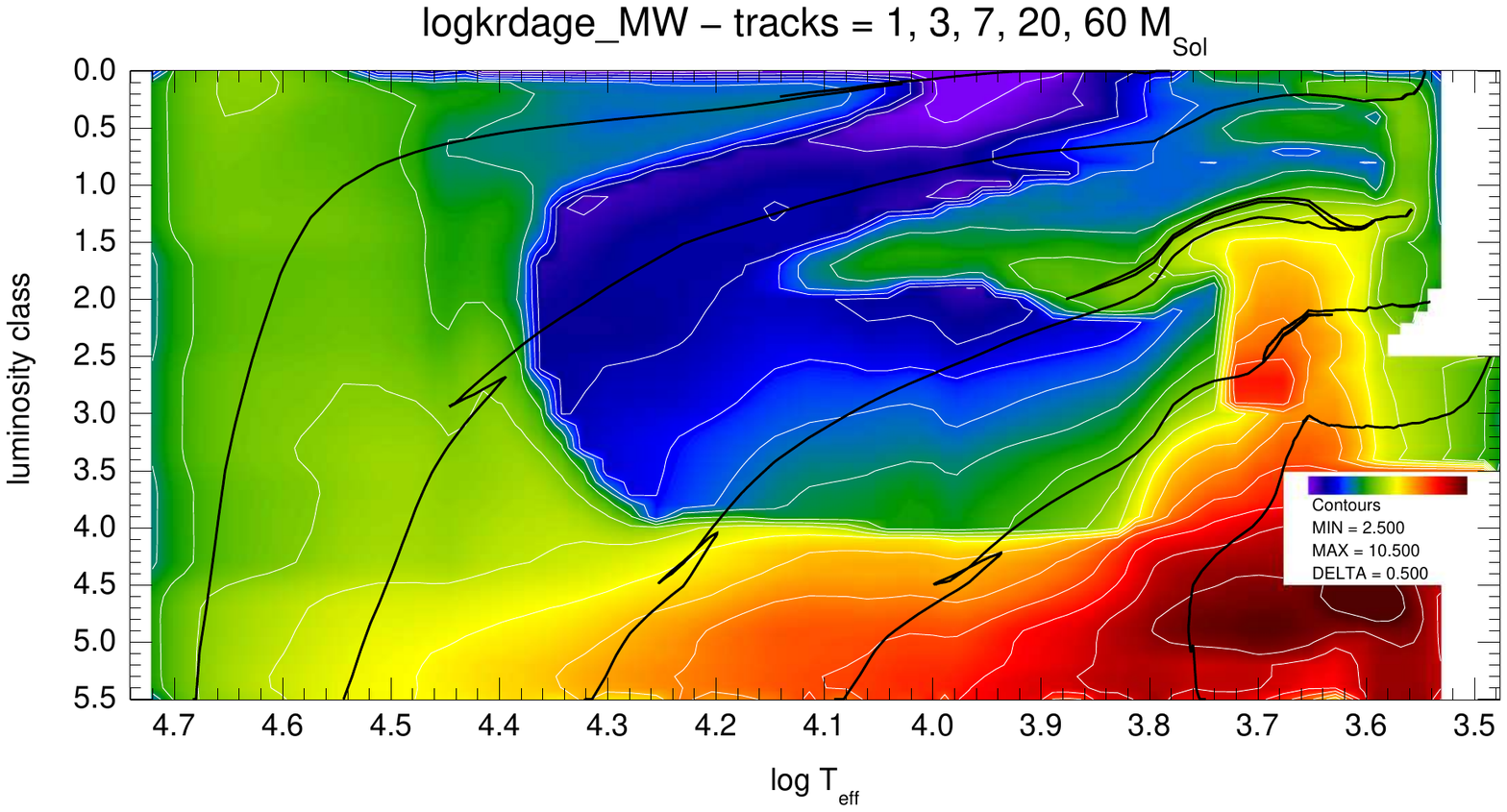}
}
\caption{Mean age (top), time spent in a cell (center), and IMF + age range weight (bottom) $T_{\rm eff}$ - LC grid for the MW. 
The top and center panels are shown in units of log a with minima of 5 and 2, maxima of 10 and 9, and spacings of 0.25 and 0.50, respectively.
The evolutionary tracks for $m_i$ = 1, 3, 7, 20, and 60 M$_\odot$ are also plotted.}
\end{figure}

\begin{figure}
\centerline{
\includegraphics[width=0.85\linewidth, bb=28 28 566 324]{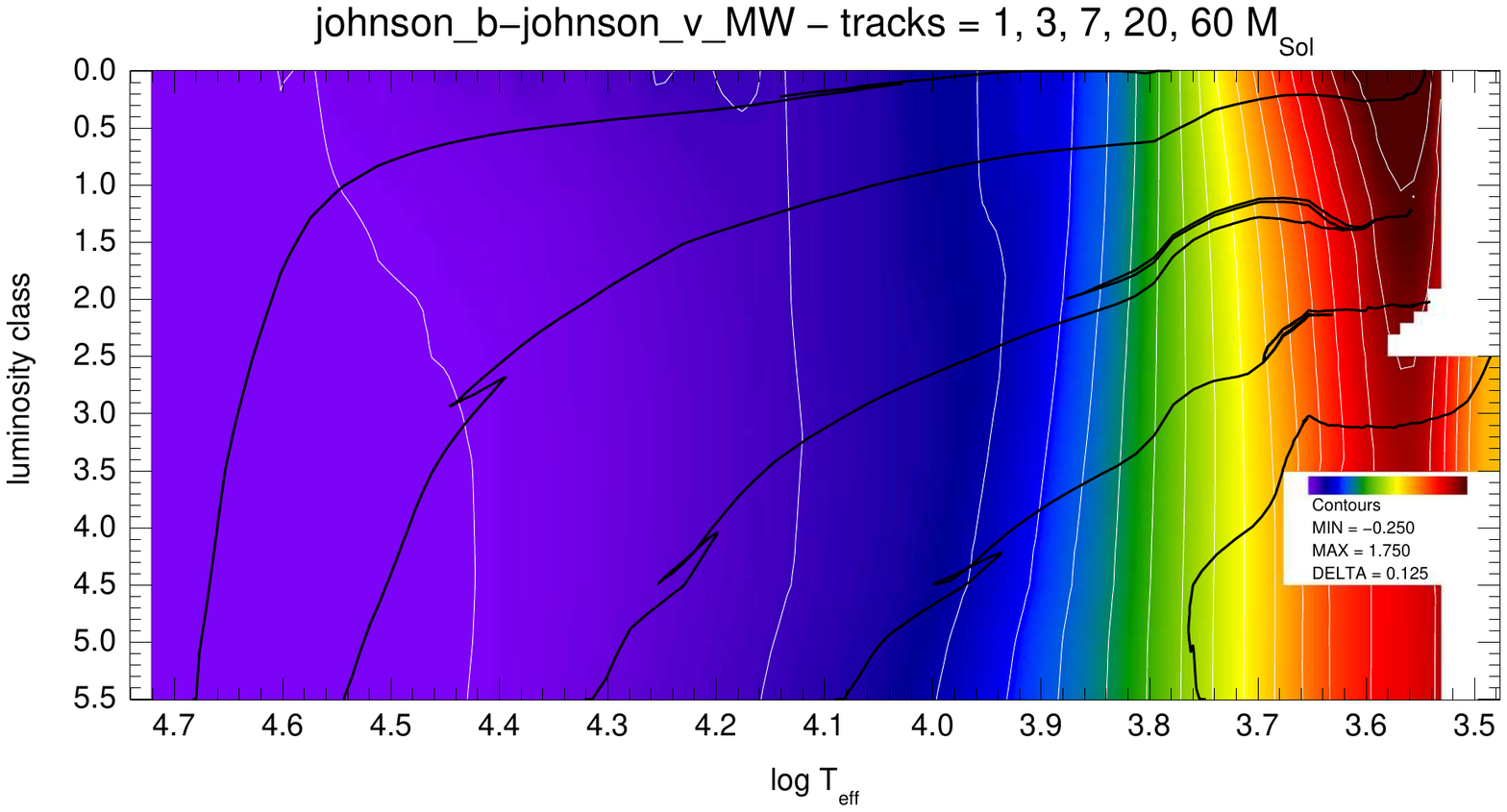}
}
\centerline{
\includegraphics[width=0.85\linewidth, bb=28 28 566 324]{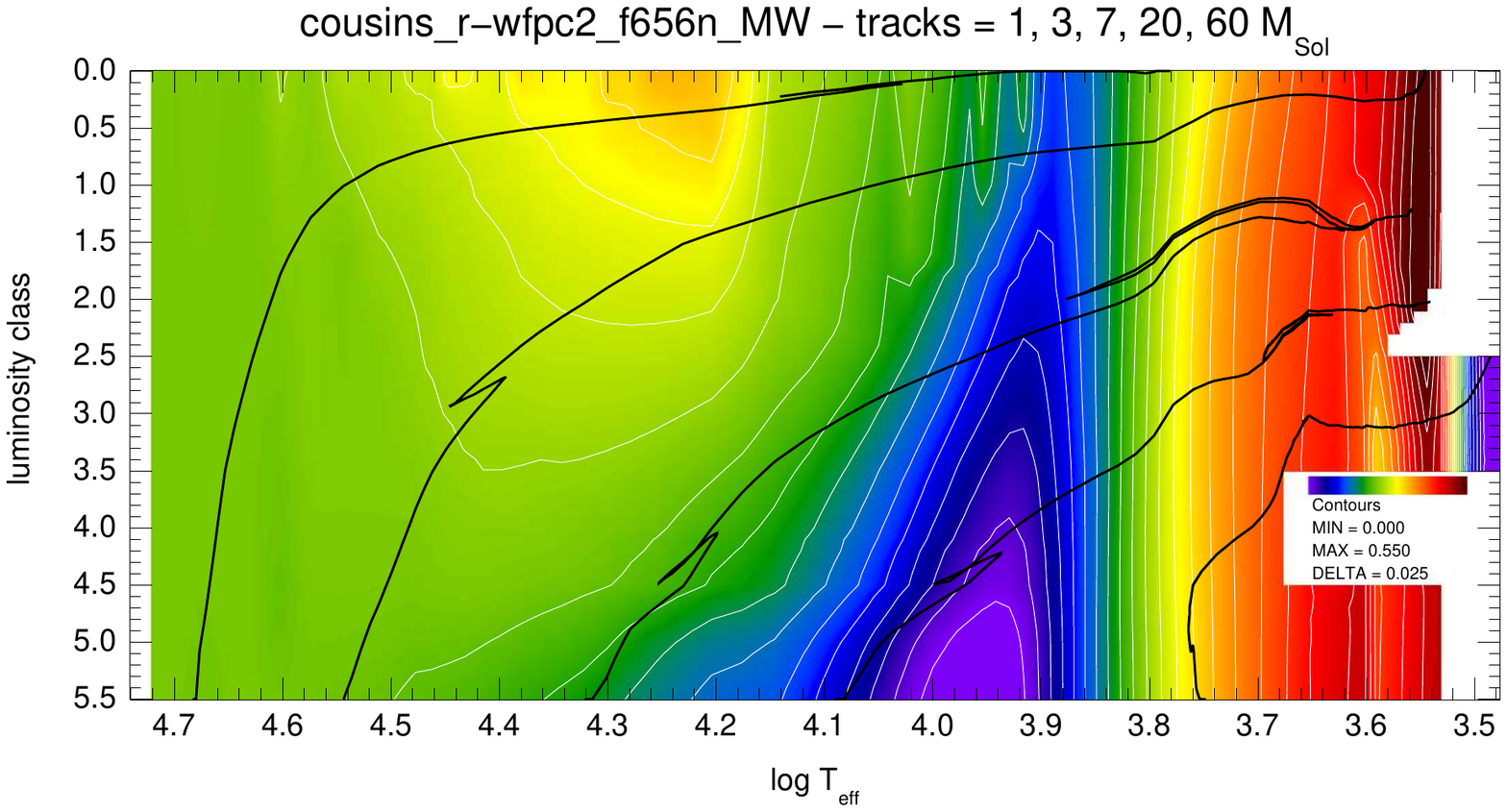}
}
\centerline{
\includegraphics[width=0.85\linewidth, bb=28 28 566 324]{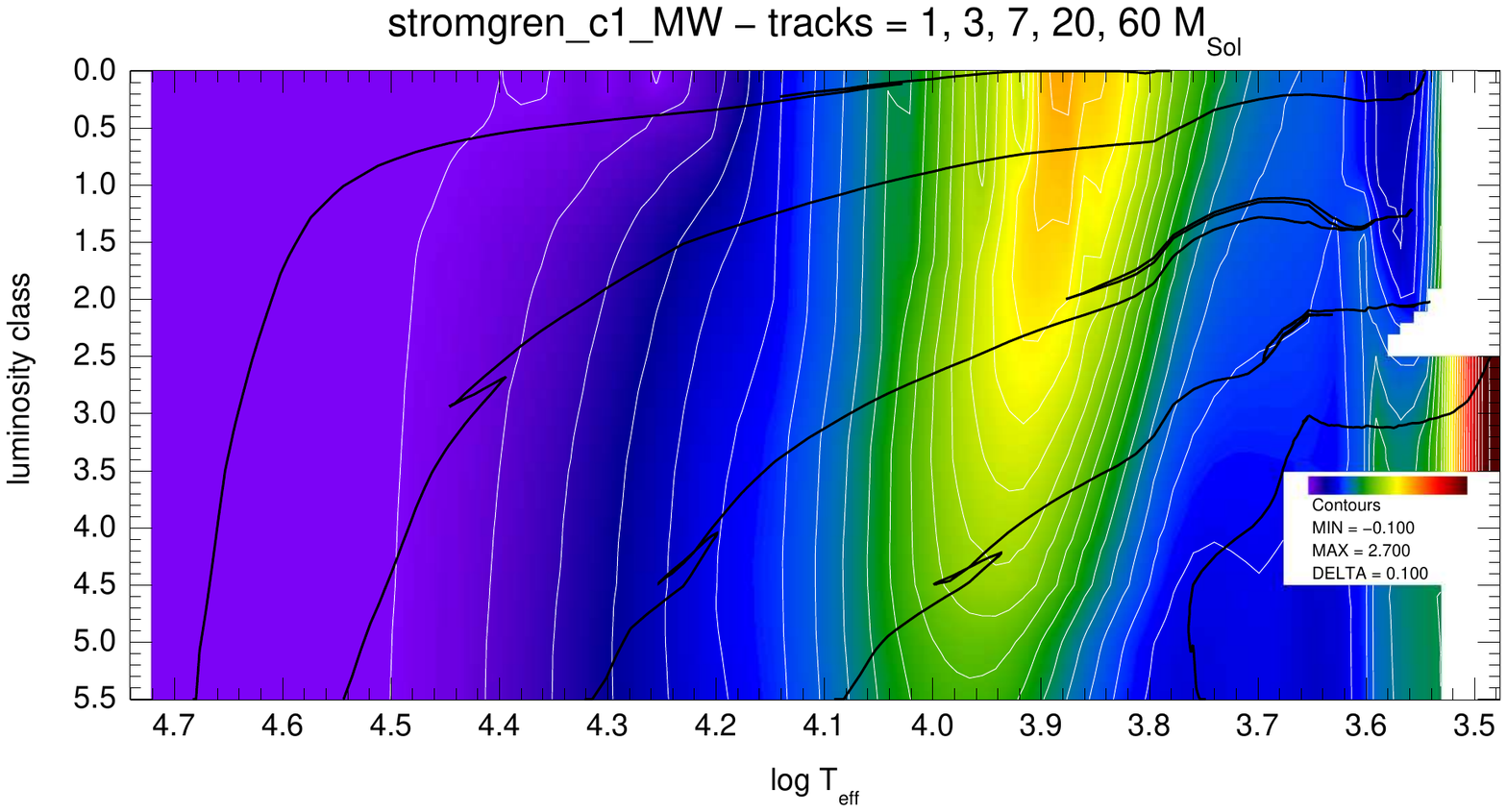}
}
\caption{Johnson $B-V$ (top), Cousins $R -$ WFPC2 F656N (center), and Str\"omgren $c_1$ (bottom) $T_{\rm eff}$ - LC grid for the MW. 
All panels are shown in magnitudes with minima of $-0.25$, $0.00$, and $-0.10$; maxima of $1.75$, $0.55$, and $2.70$; 
and spacings of $0.125$, $0.025$, and $0.100$, respectively.
The evolutionary tracks for $m_i$ = 1, 3, 7, 20, and 60 M$_\odot$ are also plotted.}
\end{figure}

\begin{figure}
\centerline{
\includegraphics[width=0.85\linewidth, bb=28 28 566 324]{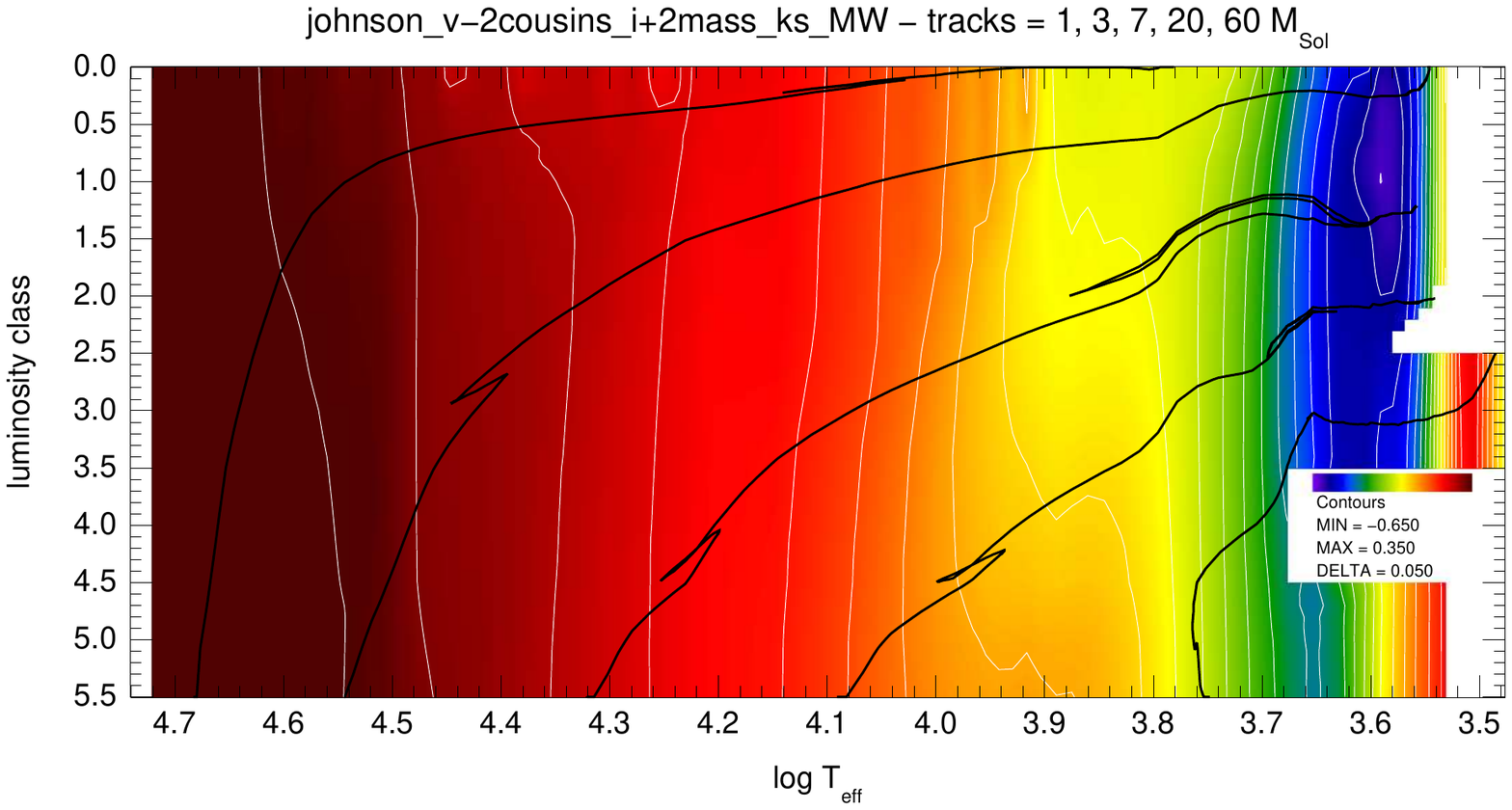}
}
\centerline{
\includegraphics[width=0.85\linewidth, bb=28 28 566 324]{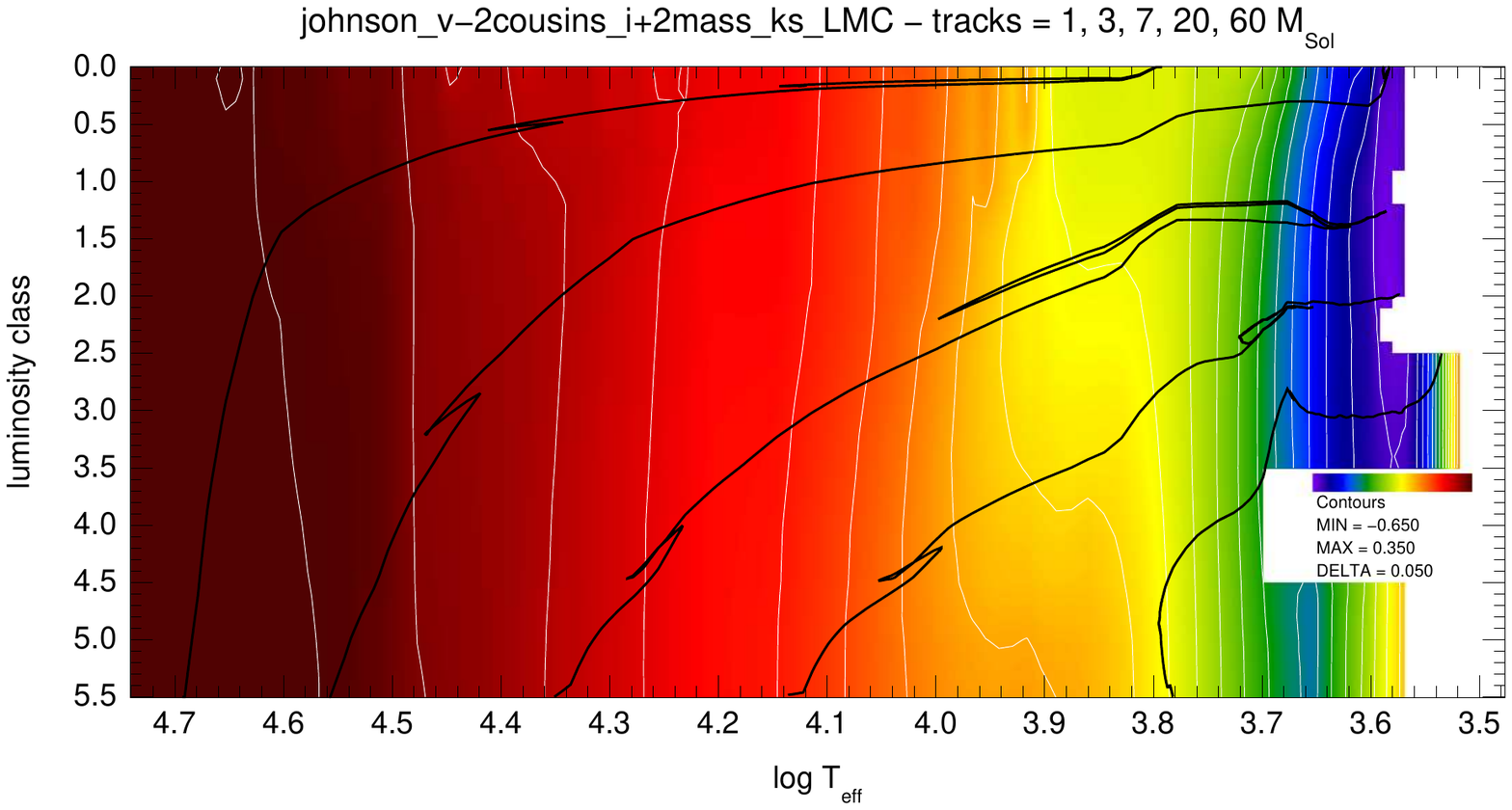}
}
\centerline{
\includegraphics[width=0.85\linewidth, bb=28 28 566 324]{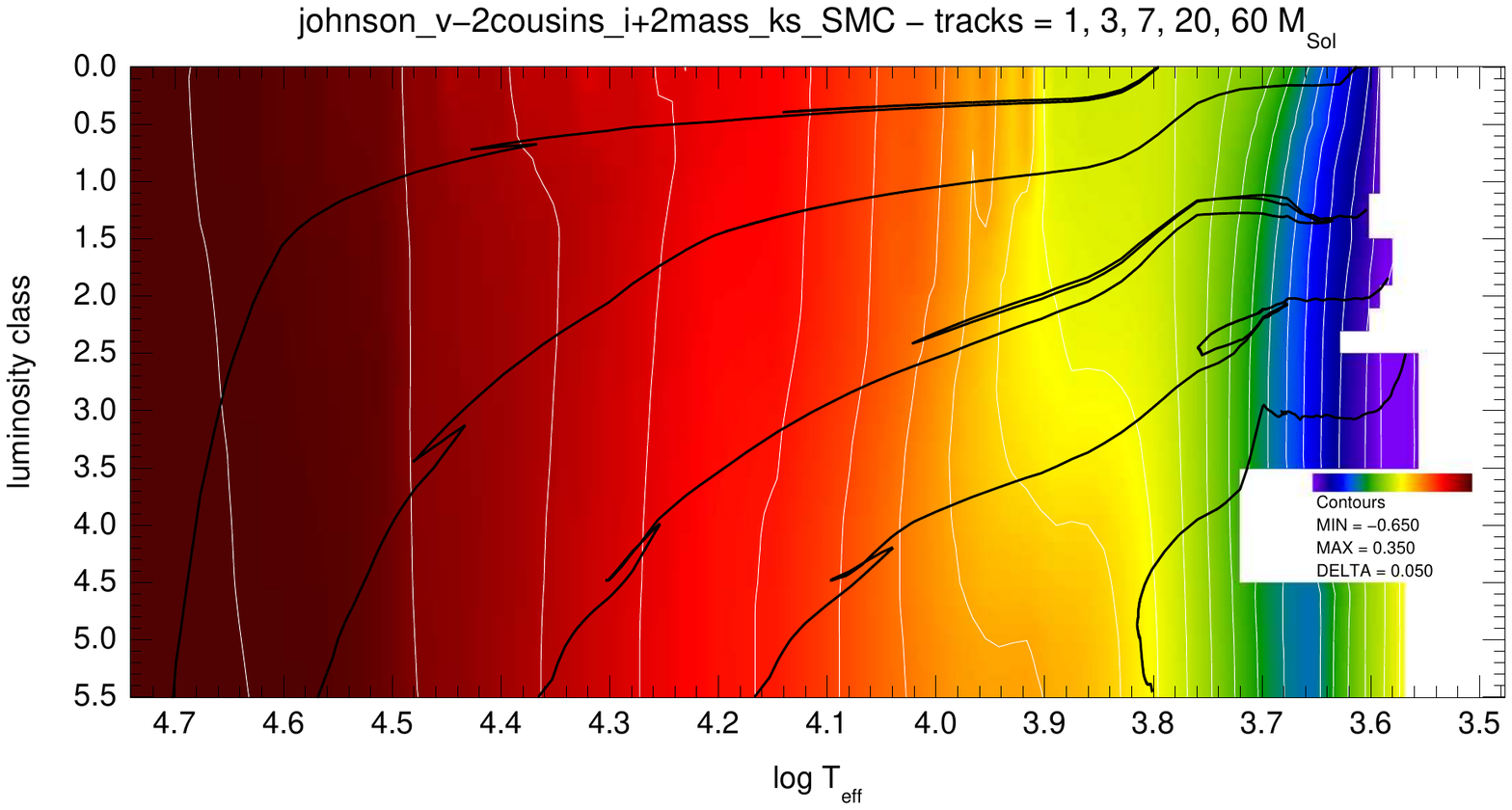}
}
\caption{Johnson-Cousins-2MASS $V-2I+K$ index for the MW (top), LMC (center), and SMC (bottom) $T_{\rm eff}$ - LC grids. 
In all cases the contours are spaced at $0.05$ magnitude intervals between $-0.65$ and $0.35$.
The evolutionary tracks for $m_i$ = 1, 3, 7, 20, and 60 M$_\odot$ are also plotted.}
\end{figure}

\small  
%
%

%
%
%
%
%
\bibliographystyle{aj}
\small
\bibliography{general}

\end{document}